\documentclass[journal]{IEEEtran}

\usepackage{amsmath}
\usepackage{suffix}
\usepackage{mathtools}
\usepackage{cite}
\usepackage{color,soul}
 
\usepackage{multirow}
\usepackage{hhline}

\ifCLASSINFOpdf
\else
\fi

\usepackage{suffix,mathtools}
\DeclarePairedDelimiterX\MeijerM[3]{\lparen}{\rparen}%
{#3\,\delimsize\vert\begin{smallmatrix}#1 \\ #2\end{smallmatrix}}

\newcommand\MeijerG[8][]{%
  G^{\,#2,#3}_{#4,#5}\MeijerM[#1]{#6}{#7}{#8}}

\WithSuffix\newcommand\MeijerG*[7]{%
  G^{\,#1,#2}_{#3,#4}\MeijerM*{#5}{#6}{#7}}

\ifCLASSINFOpdf
\else
\fi

\begin{document}

\title{Exact Outage Probability Analysis  of the Mixed  RF/FSO System with Variable Gain Relays}

\author{Milica~I.~Petkovic,~\IEEEmembership{Member,~IEEE},~Zeljen Trpovski,~\IEEEmembership{Member,~IEEE}
\thanks{M.~I.~Petkovic and Z.~Trpovski are with University of Novi Sad, Faculty of Technical Sciences, Trg Dositeja Obradovica 6, 21000 Novi Sad, Serbia (e-mails: milica.petkovic@uns.ac.rs; zeljen@uns.ac.rs).}
}


\maketitle

\begin{abstract}
This paper presents an unified  analysis of the mixed radio-frequency/free-space optics (RF/FSO) relaying system, with multiple variable gain  amplify-and-forward relays. The partial relay selection (PRS) is employed to select the active relay for further re-transmission. Due to fast fading statistics of the first RF hop, it is assumed that the channel state information of the RF link is outdated, which is used for both relay gain adjustment and PRS procedure. The RF hops are subject to the Rayleigh fading, while the FSO hop is affected by the atmospheric turbulence and the pointing errors. The intensity fluctuations of the optical signal caused by atmospheric turbulence are modeled by general M$\acute{{\rm{a}}}$laga ($ \mathcal{M} $) distribution, which takes into account effect of multiple scattered components.
Exact expression for the outage probability is derived. In addition, high signal-to-noise ratio approximations are provided, which can be used to efficiently determine the outage probability floor. Numerical results are validated by Monte Carlo simulations, which are used to examine the effects of the system  and channel parameters on the RF/FSO system performance.
\end{abstract}


\IEEEpeerreviewmaketitle

\section{Introduction}
\IEEEPARstart{I}{n} order to ensure and support demanding multimedia services and applications in the future $ 5^{th} $ generation wireless networks, introducing and/or combining already existing communication technologies with the novel ones is required. Since the optical fiber implementation is proved to be quite complicated and expensive in some areas, research interest in optical wireless technology has become renewed due to many useful benefits. Free-space optics (FSO) represents outdoor link, which uses infrared band and  provides high bandwidth capacity and operation in licence-free unregulated spectrum, with very easy and low-cost implementation and repositioning possibility \cite{book, survey, new1, new2}. 

The FSO signal transmission via atmospheric channel is seriously affected by few phenomena, such as the atmospheric turbulence and the misalignment between FSO transmitter and receiver (pointing errors) \cite{book, survey, PE2, PE4}. Furthermore, aggravating requirement for the FSO system application, is obligatory line-of-sight (LOS) existence between FSO transmitter and receiver. Some environment scenarios, such as difficult terrains and crowded urban streets, make very hard or even impossible to provide LOS component in wide areas. For that reason, utilization of the relaying technology within FSO systems has been proposed to realize coverage area extension where LOS cannot be achieved. In \cite{lee}, a mixed dual-hop amplify-and-forward (AF) relaying system, which is  composed of radio frequency (RF) and FSO links, has been proposed, providing a convenient way to enable multiple RF users to be multiplexed via a single FSO link, and to simultaneously use FSO as a last mile access network. The performance of the mixed RF/FSO system with fixed AF gain relay was analyzed in \cite{lee, endend, Ansari-Impact, Anees2, Zhang_JLT, Zedini_PhotonJ}, while the RF/FSO system with variable AF gain relay was observed in \cite{nova,Zedini_PhotonJ, JSAC1, JSAC2,var2,var3}.

In order to ensure further improvement of the system performance, implementation of multiple relays within FSO systems has been also investigated in \cite{FSO1,FSO4,FSO5}. The idea of  utilization of  multiple relays within mixed RF/FSO system has been proposed in \cite{JLT}. The study in \cite{JLT} was concentrated on the RF/FSO system with multiple fixed gain AF relays, while partial relay selection (PRS) procedure was employed to choose active relay for further transmission. The PRS technique, firstly presented in \cite{PRS}, represents the effective and low-cost technique since relay selection is proceeded based on single-hop instantaneous channel state information (CSI), avoiding additional network delays and achieving power save. In \cite{JLT}, the outage probability expressions were derived, assuming the first RF hops experience Rayleigh fading, and the  second FSO hops are influenced by Gamma-Gamma (GG) atmospheric turbulence. Further, the same RF/FSO system with fixed gain AF relays was analyzed in \cite{chapter}, while the outage probability expression was provided for the case when the pointing errors effect was taken into account. Additionally, \cite{prsF1,prsF2} considered RF/FSO system with fixed gain AF relays taking account aggregate hardware impairments.

In contrast to \cite{JLT,chapter,prsF1,prsF2}, which assumed fixed AF gain relays, this paper analyses the PRS based multiple RF/FSO system with variable AF gain relays. The main contribution of the paper is to provide exact expression for the outage probability, considering that 
the optical signal intensity fluctuations due to atmospheric turbulence are modeled by general  M$\acute{{\rm{a}}}$laga ($ \mathcal{M} $) distribution, which takes into account multiple scattering effects and represents more general model compared to GG distribution  \cite{JSAC1,FSO5,M1,M4}. In addition, the pointing errors are taken into account \cite{endend,M5,M6,M7}. 
Assuming that the RF signal transmission from source to the relay station is performed in frequency range from 900 MHz to 2.4 GHz, due  to fast fading statistic, the estimation of the RF channel is happened to be imperfect, so the \textit{outdated} CSI  is used for both relay selection and relay gain regulation.
The first RF hops are introduced since the LOS is not provided in that area, so the RF links are assumed to be subject to Rayleigh fading.
In practical system scenario, it can be possible that the relay with best estimated CSI is not able to forward the signal. This problem is also taken into consideration in \cite{prs1}. 
A novel outage probability expression is derived, which is further simplified to  some special cases. 
Approximate outage probability expressions are also provided, which are utilized to determine the outage probability floor.
Furthermore, as a special case when only one relay is assumed, the
outage probability expression is simplified
to the corresponding results already reported in \cite{JSAC2}. 
Based on derived analytical expressions, numerical results are obtained and validated by Monte Carlo simulations, which represent widely used computing algorithms performed  to obtain and confirm numerical results by generating repeated random sampling.

The rest of the paper is organized as follows. Section II presents
the system and channel model. The outage probability analysis is described in Section III, which also contains some special cases and approximation expressions. Numerical results and simulations with discussions are given in Section IV. Some concluding remarks are presented in Section V.

\section{System and channel model}

\begin{figure}[!b]
\centering
\includegraphics[width=3.4in]{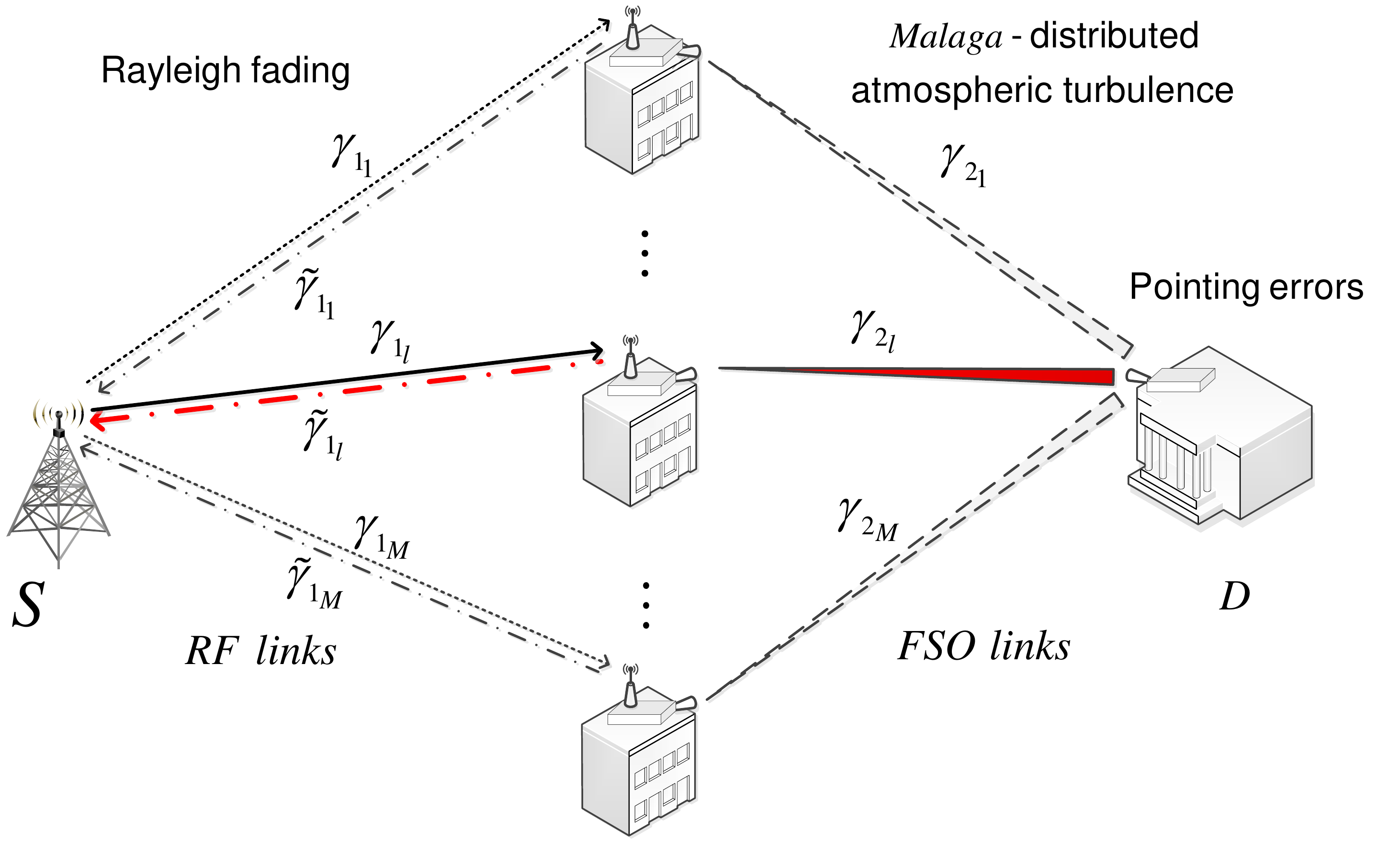}
\caption{Multiple mixed  RF/FSO system based on PRS with outdated CSI.}
\label{Fig_1}
\end{figure}

Fig.~\ref{Fig_1} presents the mixed RF/FSO system with multiple variable gain relays. The observed system consists of a source, $S$, a destination, $D$, and $M \ge 1$ relays. The node $S$ continuously monitors and  periodically estimates the conditions of the $ S-R$ RF channels. The node $S$ selects the active relay $ R_l $, which is the one with best estimated CSI of the hop between source and relay. In the case the best selected relay is not able to perform further communication, next best relay is chosen, etc. In other words, the $ l $th worst or $ (M-l) $th  best relay is selected \cite{prs1}.

The RF hops are considered in the first part of the system since the LOS component is not provided in that area. For that reason, Rayleigh distribution is assumed to describe the RF channel fading conditions. In practice, temporal variations of the RF channel occurs. Hence, the errors in channel estimation can happen, and the estimated CSI used for relay selection is not the same as the one at the time when transmission is performed. It means that the channel estimation at the relay is imperfect, and PRS is performed based on outdated CSI. 
Since variable  relays are employed, the gain is determined based on short-term statistics of the RF hops. In order to avoid additional channel estimation, the outdated CSI used for relay selection  is also utilized to adjust relay gain.

In the first phase of the transmission, after selection of the active relay, $ R_l $, signal is transmitted over RF link. Received electrical signal at the relay $ R_l $ is defined as 
\begin{equation}
{r_{R_l}} = {h_{SR_l}}r + {n_{SR}},
\label{signal_r}
\end{equation}
where $ r $ denotes the  signal with an average power $ P_s $, sent from the node $ S $. The fading amplitude over the RF link is denoted by $ h_{SR_l} $, with $ {\rm E}[ {h_{SR_l}^2}] = 1 $  $ ({\rm E}[\cdot] $ is mathematical expectation$)$. An additive white Gaussian noise (AWGN) with zero mean and variance  $ \sigma_{SR}^2$ is denoted by $n_{SR} $.
Signal at the relay is amplified by  gain $ G $ based on outdated CSI, which is defined as \cite{JSAC2, prs3}
\begin{equation}
{G^2} = \frac{1}{{\tilde h_{SR_l}^2{P_s}}},
\label{gain}
\end{equation}
where $ \tilde h_{SR_l} $ represents the estimated version of $ h_{SR_l} $.

In the next phase, amplified signal is converted to the optical one by subcarrier intensity modulation. DC bias is added to satisfy the non-negativity constraint. The optical signal at the relay is defined as
\begin{equation}
{r_{o}} ={P_t} \left( {1 + mG{r_{R_l}}} \right),
\label{signal_ro}
\end{equation}
where $ P_t $ represents the average transmitted optical power, $ m $ denotes  modulation index $(m=1)$, and $ r_{R_l} $ is defined in (\ref{signal_r}). Optical signal is further forwarded to the destination via atmospheric turbulence channel. At the destination, direct detection is done, DC bias is removed, and an optical-to-electrical conversion is performed via PIN photodetector. The received electrical signal is  expressed as
\begin{equation}
\begin{split}
{r_D}&= {I_{{R_l}D}}\eta{P_t}G r_{R_l} + {n_{RD}} \\
&= {I_{{R_l}D}}\eta{P_t}G \left( {{h_{SR}}_lr + {n_{SR}}} \right) + {n_{RD}},
\end{split}
\label{signal_rd}
\end{equation}
where $ I_{R_lD} $ represents the intensity of an optical signal, and $ \eta $ denotes an optical-to-electrical conversion coefficient. The AWGN over FSO link with zero mean and variance $ \sigma _{RD}^2 $ is denoted by $ n_{RD} $.

Following (\ref{gain}) and (\ref{signal_rd}), the equivalent instantaneous overall signal-to-noise ratio (SNR) at the  destination is defined as 
\begin{equation}
{\gamma _{eq}} = \frac{I_{{R_l}D}^2{\eta ^2}{P_t^2{G^2}h_{S{R_l}}^2{P_s}}}{{I_{{R_l}D}^2{\eta ^2}P_t^2{G^2}\sigma _{SR}^2 + \sigma _{RD}^2}} = \frac{{{\gamma _{1_l}}{\gamma _{2_l}}}}{{{\gamma _{2_l}} + {{\tilde \gamma }_{1_l}}}},
\label{eq_snr}
\end{equation}
where  ${\gamma _{{1_l}}} = {{h_{S{R_l}}^2{P_s}} \mathord{\left/
 {\vphantom {{h_{S{R_l}}^2{P_s}} {\sigma _{SR}^2}}} \right.
 \kern-\nulldelimiterspace} {\sigma _{SR}^2}} = \;h_{S{R_l}}^2{\mu _1} $ represents the instantaneous SNR of the first RF hop with the average SNR defined as $ {\mu _1} = {\rm{E}}\left[ {{\gamma _{{1_l}}}} \right] = {{{P_s}} \mathord{\left/
 {\vphantom {{{P_s}} {\sigma _{SR}^2}}} \right.
 \kern-\nulldelimiterspace} {\sigma _{SR}^2}} $; $ {\tilde \gamma _{{1_l}}} = {{\tilde h_{S{R_l}}^2{P_s}} \mathord{\left/
 {\vphantom {{\tilde h_{S{R_l}}^2{P_s}} {\sigma _{SR}^2}}} \right.
 \kern-\nulldelimiterspace} {\sigma _{SR}^2}}$ is its estimated version; and the instantaneous SNR of the  FSO hop is defined as $ {\gamma _{{2_l}}} = {{I_{{R_l}D}^2{\eta ^2}P_t^2} \mathord{\left/
 {\vphantom {{I_{{R_l}D}^2{\eta ^2}P_t^2} {\sigma _{RD}^2}}} \right.
 \kern-\nulldelimiterspace} {\sigma _{RD}^2}} $. The electrical SNR is defined as $ {\mu _2} = {{{{\rm{E}}^2}\left[ {{I_{{R_l}D}}} \right]{\eta ^2}P_t^2} \mathord{\left/
 {\vphantom {{{{\rm{E}}^2}\left[ {{I_{{R_l}D}}} \right]{\eta ^2}P_t^2} {\sigma _{RD}^2}}} \right.
 \kern-\nulldelimiterspace} {\sigma _{RD}^2}} $.

\subsection{RF channel model}

Since the RF hops experience Rayleigh fading, the instantaneous SNR of the first RF hop and its estimated version are two exponentially distributed correlated RVs. Since PRS with outdated CSI is considered, as well as possibility that the best relay is not able to perform transmission and the $ l $th worst (or $ (M-l) $th  best) relay is selected, the  joint probability density function (PDF) of SNRs ${\gamma _{{1_l}}}$ and $ {\tilde\gamma _{{1_l}}}$ is expressed as \cite[(48)]{prs3}
\begin{equation}
\begin{split}
{f_{{\gamma _{1_l}},{{\tilde \gamma }_{1_l}}}}\left( {x,y} \right) & = l{M \choose l}\frac{{e^{ - \frac{x}{{\left( {1 - \rho } \right){\mu _1}}}}}}{{\left( {1 - \rho } \right)\mu _1^2}}{I_0}\left( {\frac{{2\sqrt {\rho xy} }}{{\left( {1 - \rho } \right){\mu _1}}}} \right)\\
& \times \sum\limits_{i = 0}^{l - 1}\! {l-1 \choose i} {\left( { - 1} \right)^i}{e^{ - \frac{{\psi_i y}}{{\left( {1 - \rho } \right){\mu _1}}}}},
\end{split}
\label{pdf_rf}
\end{equation}
where $ \rho $ is correlation coefficient, $\psi_i  = \left( {M - l + i} \right)\left( {1 - \rho } \right) + 1 $, and $ {I_\nu} \left(  \cdot  \right)$ represents the $ \nu $th order modified Bessel function of the first kind \cite[(8.406)]{Grad}.

\subsection{FSO channel model}

The intensity fluctuations of the received optical signal, caused by atmospheric turbulence, are modeled by recently presented M$\acute{{\rm{a}}}$laga ($ \mathcal{M} $) distribution \cite{M1,M4,M5,M6,M7}. The pertinence  of the $ \mathcal{M} $ distribution in regard to the ones earlier considered in literature (such as Log-normal, K, GG, Exponential, etc.), is the consideration of the multiple scattering effects \cite{M1}.  More precisely, presented model includes three components. Beside an one, $ U_L $, which occurs due to LOS contribution, there are two more components occurred due to scattering effects:  the component which is scattered by the eddies on the propagation axis, $ U_S^{C} $, and the one scattered by the off-axis eddies, $ U_S^{G} $. The component $ U_S^{C} $ is coupled to $ U_L $, while the component $ U_S^{G} $ is statistically independent from both $ U_L $ and $ U_S^{C} $. More details about $ \mathcal{M} $ distribution can be found in \cite{M1}. In addition, pointing errors are taken into consideration. The PDF of the optical signal intensity is derived as \cite[(5)]{M7} 
\begin{equation}
\begin{split}
{f_{{I_{{R_l}D}}}}\left( I \right)& = \frac{{{\xi ^2}{\rm A}}}{2}{I^{ - 1}}\sum\limits_{k = 1}^\beta  {{a_k}} {\left( {\frac{{\alpha \beta }}{{g\beta  + \Omega '}}} \right)^{ - \frac{{\alpha  + k}}{2}}} \\
&\times \MeijerG*{3}{0}{1}{3}{\xi^2+1}{\xi ^2, \, \alpha, \, k}{{\frac{{\alpha \beta }}{{g\beta  + \Omega '}}\frac{I}{{{A_0}{I_l}}}}}, 
\end{split}
\label{pdf_I}
\end{equation}
where $ \beta $ is natural number representing the amount of fading parameter,  $ G_{p,q}^{m,n}\left(  \cdot  \right) $ is Meijer's \textit{G}-function \cite[(9.301)]{Grad}, and constants $ \rm A $ and $ a_k $ are defined as \cite[(25)]{M1}
\begin{equation}
{\rm A} \buildrel \Delta \over = \frac{{2{\alpha ^{\frac{\alpha }{2}}}}}{{{g^{1 + \frac{\alpha }{2}}}\Gamma \left( \alpha  \right)}}{\left( {\frac{{g\beta }}{{g\beta  + \Omega '}}} \right)^{\beta  + \frac{\alpha }{2}}},
\label{const1}
\end{equation}
\begin{equation}
{a_k} \buildrel \Delta \over = {\beta-1 \choose k-1}
\frac{{{{\left( {g\beta  + \Omega '} \right)}^{1 - \frac{k}{2}}}}}{{\left( {k - 1} \right)!}}{\left( {\frac{{\Omega '}}{g}} \right)^{k - 1}}{\left( {\frac{\alpha }{\beta }} \right)^{\frac{k}{2}}},
\label{const2}
\end{equation} 
with a positive parameter $\alpha $ related to the effective number of large-scale cells of the scattering process. Further, 
$g~=~{\rm E}\left[ {{{\left| {U_S^G} \right|}^2}} \right] = 2{b_0}\left( {1 - \rho_M } \right) $ represents the average power of the scattering component received by off-axis eddies, where $ 2{b_0} = {\rm E}\left[ {{{\left| {U_S^C} \right|}^2} + {{\left| {U_S^G} \right|}^2}} \right] $ defines the average power of the total scatter components. The  amount of scattering power coupled to the LOS component is denoted by $ 0~\le~\rho_M~\le 1$. The average power from the coherent contributions is expressed as $ {\Omega'} = \Omega  + 2{b_0}\rho +  2\sqrt {2{b_0}\rho_M \Omega } \cos \left( {{\phi _A} - {\phi _B}} \right) $, where $ \Omega  = {\rm E}\left[ {{{\left| {{U_L}} \right|}^2}} \right] $ represents the average power of the LOS component. Deterministic phases of the LOS and the coupled-to-LOS scatter terms are denoted as $ \phi _A$ and $ \phi _B $, respectively.

The path loss component is  defined by deterministic model as  $  {I_l}=~\exp \left( { - \chi d} \right) $ \cite{PE2}, where $ \chi $ denotes the atmospheric attenuation coefficient and $ d $ represents length of the FSO link. The parameter $ {\xi} $ is defined as
\begin{equation}
\xi  = \frac{{{a_{{d_{eq}}}}}}{{2{\sigma _s}}},
 \label{ksi}
\end{equation}
with  the equivalent beam radius at the receiver  and the pointing error (jitter) standard deviation at the receiver denoted  by  $ {a_{{d_{eq}}}} $ and $ \sigma_s $, respectively. 
Further, the parameter $ {a_{{d_{eq}}}} $  is related to  the beam radius at the distance $ d $, $ a_d $, as  $a_{{d_{eq}}}^2=~a_d^2\sqrt \pi {{\operatorname{erf} (v)} \mathord{\left/
 {\vphantom {{erf(v)} {(2v\exp ( - {v^2}))}}} \right.
 \kern-\nulldelimiterspace} {(2v\exp ( - {v^2}))}} $, with $ v =~{{\sqrt \pi  a} \mathord{\left/
 {\vphantom {{\sqrt \pi  a} {(\sqrt 2 {a_d})}}} \right.
 \kern-\nulldelimiterspace} {(\sqrt 2 {a_d})}} $, and the parameter $ a$ denotes the radius of a circular detector aperture. 
The parameter $ A_0 $ is defined as  $ {A_0} = {\left[ {\operatorname{erf} \left( v \right)} \right]^2} $, where $ \operatorname{erf} \left(  \cdot  \right) $ is the error function \cite[(8.250.1)]{Grad}. Next, the parameter $ a_d $ is dependent on the optical beam radius  at the waist, $ a_0 $, and to the radius of curvature, $ F_0 $, as  ${a_d}\!=~\!{a_0}{\left( {({\Theta _o} + {\Lambda _o})(1 + 1.63\sigma_R^{12/5}{\Lambda _1})} \right)^{1/2}}$, where $ {\Theta _o} =1-{d \mathord{\left/
 {\vphantom {L {{F_0}}}} \right.
 \kern-\nulldelimiterspace} {{F_0}}}$, $ {\Lambda _o} = {{2d} \mathord{\left/
 {\vphantom {{2d} {(\iota a_0^2)}}} \right.
 \kern-\nulldelimiterspace} {(\iota a_0^2)}}$,  $ {\Lambda _1} = {{{\Lambda _o}} \mathord{\left/
 {\vphantom {{{\Lambda _o}} {(\Theta _o^2 + \Lambda _o^2)}}} \right.
 \kern-\nulldelimiterspace} {(\Theta _o^2 + \Lambda _o^2)}} $ \cite{PE4}. The Rytov variance determines the optical signal intensity due to atmospheric turbulence. It is defined as  $ \sigma_R^{2}=1.23C_n^{2}\iota^{7/6}d^{11/6} $, where $ \iota = 2\pi/\lambda $ is the wave number with the wavelength $ \lambda $, and $ C_n^{2} $ is the refractive index structure parameter.

Based on definition of the instantaneous SNR of FSO hop, $ \gamma_{2_l} $, and the PDF of  $ I_{R_lD} $ in (\ref{pdf_I}), after some mathematical manipulations, the PDF of $ \gamma_{2_l} $ is easily derived as \cite[(7)]{M7}
\begin{equation}
\begin{split}
{f_{{\gamma _{2}}}}\left( \gamma  \right)& = \frac{{{\xi ^2}{\rm A}}}{{4\gamma }}\sum\limits_{k = 1}^\beta  {{a_k}} {\left( {\frac{{\alpha \beta }}{{g\beta  + \Omega '}}} \right)^{ - \frac{{\alpha  + k}}{2}}} \\
&\times \MeijerG*{3}{0}{1}{3}{\xi^2+1}{\xi ^2, \, \alpha, \, k}{\frac{{\alpha \beta \kappa \left( {g + \Omega '} \right)}}{{g\beta  + \Omega '}}\sqrt {\frac{\gamma }{{{\mu _2}}}} }, 
\end{split}
\label{pdf_g2}
\end{equation}
where $ \kappa  = {{{\xi ^2}} \mathord{\left/
 {\vphantom {{{\xi ^2}} {\left( {{\xi ^2} + 1} \right)}}} \right.
 \kern-\nulldelimiterspace} {\left( {{\xi ^2} + 1} \right)}} $. Based on definition of the moments of the combined model ($ \mathcal{M} $ distribution + pointing errors) presented in \cite[(33)]{M5}, the electrical SNR is determined as ${\mu _2} = {{{\eta ^2}P_t^2{A_0}^2{I_l}^2{\kappa ^2}{{\left( {g + \Omega '} \right)}^2}} \mathord{\left/
 {\vphantom {{{\eta ^2}P_t^2{A_0}^2{I_l}^2{\kappa ^2}{{\left( {g + \Omega '} \right)}^2}} {\sigma _{RD}^2}}} \right.
 \kern-\nulldelimiterspace} {\sigma _{RD}^2}} $ \cite{M7}.
The cumulative distribution function (CDF) of $ \gamma_{2_l} $ is obtained as
\begin{equation}
\begin{split}
{F_{{\gamma _{2}}}}\left( \gamma  \right)& =\! \!\int\limits_0^\gamma \!\! {{f_{{\gamma _{2}}}}\left( x \right)} dx = \frac{{{\xi ^2}{\rm A}}}{2}\sum\limits_{k = 1}^\beta  {{a_k}} {\left( {\frac{{\alpha \beta }}{{g\beta  + \Omega '}}} \right)^{ - \frac{{\alpha  + k}}{2}}} \\
&\times \MeijerG*{3}{1}{2}{4}{1, \, \xi^2+1}{\xi ^2, \, \alpha, \, k, \, 0}{\frac{{\alpha \beta \kappa \left( {g + \Omega '} \right)}}{{g\beta  + \Omega '}}\sqrt {\frac{\gamma }{{{\mu _2}}}}}. 
\end{split}
\label{cpdf_g2}
\end{equation}

\section{Outage probability analysis}

As one of the most important system performance metric, the outage probability indicates how often the system is under the desired performance threshold.
The outage probability, defined as the probability that the instantaneous overall SNR, $\gamma_{eq}$, defined in (\ref{eq_snr}), falls below a predetermined outage threshold, $\gamma_{th}$, is given by 
\begin{equation}
\begin{split}
{P_{out}} = {F_{eq}}\left( {{\gamma _{th}}} \right) = \Pr \left( \gamma_{eq}< {\gamma _{th}} \right),
\end{split}
\label{pout01}
\end{equation}
where $\Pr\left(  \cdot  \right)$ denotes probability. After substituting (\ref{eq_snr})  into (\ref{pout01})  and applying some mathematical manipulations, (\ref{pout01}) is re-written as
\begin{equation}
\begin{split}
&{P_{out}} = \Pr \left( {\frac{{{\gamma _{{1_l}}}{\gamma _{{2_l}}}}}{{{\gamma _{{2_l}}} + {{\tilde \gamma }_{{1_l}}}}} < {\gamma _{th}}\left| {{\gamma _{{1_l}}},{{\tilde \gamma }_{{1_l}}}} \right.} \right) \\
&= \!\!1 - \!\!\!\int\limits_0^\infty \!\! {\int\limits_0^\infty  \!\!{\Pr \left(\! {{\gamma _{{2_l}}} > \frac{{{\gamma _{th}}y}}{x}} \right)\!\!{f_{{\gamma _{1_l}},{{\tilde \gamma }_{1_l}}}}\left( {x\!+\!{\gamma _{th}},y} \right)dxdy} } \\
& =\!\! 1 -\!\!\! \int\limits_0^\infty  \!\!{\int\limits_0^\infty  {{{\bar F}_{{\gamma _2}}}\left( {\frac{{{\gamma _{th}}y}}{x}} \right){f_{{\gamma _{1_l}},{{\tilde \gamma }_{1_l}}}}\left( {x\!+\!{\gamma _{th}},y} \right)dxdy} },
\end{split}
\label{pout2}
\end{equation}
where ${\bar F_{{\gamma _2}}}\left(  \cdot  \right) = 1 - {F_{{\gamma _2}}}\left(  \cdot  \right)$ is complementary CDF (CCDF).
After substituting (\ref{pdf_rf}) and (\ref{cpdf_g2}) into (\ref{pout2}), and mathematical derivation presented in Appendix~\ref{App1},  the analytical expression for outage probability is derived as
\begin{equation}
\begin{split}
&{P_{out}} = 1 - l {M \choose l}\sum\limits_{i = 0}^{l - 1}  \frac{{{{{l-1 \choose i} \left( { - 1} \right)}^i}}}{{M - l + i + 1}}{e^{ - \frac{{{\gamma _{th}}\left( {M - l + i + 1} \right)}}{{\psi_i {\mu _1}}}}}\\
& \!+ l{M \choose l}\!\!\sum\limits_{k = 1}^\beta  {\sum\limits_{i = 0}^{l - 1} {\sum\limits_{t = 0}^\infty  {\sum\limits_{d = 0}^t {\frac{{{{{l-1 \choose i}{t \choose d}\left( { - 1} \right)}^i}{\rho ^t}{\psi_i ^{ - (t + 1)}}}}{{\pi t{!^2}{{\left( {1 - \rho } \right)}^{t - d - 1}}\mu _1^{t - d}}}}} } } \\
&\! \times {\rm A}{a_k}{\left( {\frac{{\alpha \beta }}{{g\beta  + \Omega '}}} \right)^{ - \frac{{\alpha  + k}}{2}}}{\xi ^2}{2^{\alpha  + k - 4}}\gamma _{th}^{t - d}{e^{ - \frac{{{\gamma _{th}}}}{{\left( {1 - \rho } \right){\mu _1}}}}}\\
& \!\!\!\times \!\MeijerG*{6}{2}{3}{7}{1, \, -t, \, {\frac{{{\xi ^2} + 2}}{2}}}{{\frac{{{\xi ^2} }}{2}},\! \, \!{\frac{{\alpha}}{2},}\! \, \!{\frac{{\alpha+1}}{2},}\! \,\!{\frac{{k }}{2},} \!\,\!{\frac{{ k+1}}{2},}\! \,\!1+d,\!\,0}{\frac{{{\alpha ^2}{\beta ^2}{\kappa ^2}{{\left( {g\!+\!\Omega '} \right)}^2}\!\!{\gamma _{th}}}}{{16{{\left( {g\beta\!+\!\Omega '} \right)}^2}\psi_i{\mu _2}}}}\!.
\end{split}
\label{pout}
\end{equation}

\subsection{Special cases}
If the  amount of scattering power coupled to the LOS component is $ \rho_M=1 $ (i.e., the average power of the scattering component received by off-axis eddies, $ g $, is equal to zero), and the average power from the coherent contributions is expressed as $ {\Omega'} =1$, the $  \mathcal{M} $ distribution is reduced to the GG distribution. In that case, the product $ {{\rm A}{a_k}} $ is nonzero only when $ k=\beta $, and based on (\ref{const1}) and (\ref{const2}), it holds ${\rm{A}}{a_k}~=~{{2{\alpha ^{\frac{{\alpha  + \beta }}{2}}}{\beta ^{\frac{{\alpha  + \beta }}{2}}}} \mathord{\left/
 {\vphantom {{2{\alpha ^{\frac{{\alpha  + \beta }}{2}}}{\beta ^{\frac{{\alpha  + \beta }}{2}}}} {\left( {\Gamma \left( \alpha  \right)\Gamma \left( \beta  \right)} \right)}}} \right.
 \kern-\nulldelimiterspace} {\left( {\Gamma \left( \alpha  \right)\Gamma \left( \beta  \right)} \right)}}$. Hence, the outage probability in (\ref{pout}) is reduced to the one when the FSO hop is influenced by the GG atmospheric turbulence with the pointing errors, as
\begin{equation}
\begin{split}
&{P_{out}^{GG}} = 1 - l {M \choose l}\sum\limits_{i = 0}^{l - 1}   \frac{{{{{l-1 \choose i}\left( { - 1} \right)}^i}}}{{M - l + i + 1}}{e^{ - \frac{{{\gamma _{th}}\left( {M - l + i + 1} \right)}}{{\psi {\mu _1}}}}}\\
& \!+ \!\! \frac{{l{M \choose l}{\xi ^2}{2^{\alpha  + \beta  - 3}}}}{{\pi \Gamma \left( \alpha  \right)\Gamma \left( \beta  \right)}}
 \sum\limits_{i = 0}^{l - 1} {\sum\limits_{t = 0}^\infty  {\sum\limits_{d = 0}^t {\frac{{{l-1 \choose i}{t \choose d}{{\left( { - 1} \right)}^i}{\rho ^t}{\psi ^{ - (t + 1)}}}}{{ t{!^2}{{\left( {1 - \rho } \right)}^{t - d - 1}}\mu _1^{t - d}}}}} }  \\
&\! \!\times \!\!\gamma _{th}^{t - d}\!{e^{ - \frac{{{\gamma _{th}}}}{{\left( {1 \!-\!\rho } \right){\mu _1}}}}}\!\!\MeijerG*{6}{2}{3}{7}{1, \, -t, \, {\frac{{{\xi ^2} + 2}}{2}}}{{\frac{{{\xi ^2} }}{2}},\! \,\! {\frac{{\alpha}}{2},}\! \,\!{\frac{{\alpha+1}}{2},} \!\,\!{\frac{{\beta}}{2},} \!\,\!{\frac{{ \beta+1}}{2},}\! \,\!1+d,\!\,0}{\frac{{{\alpha ^2}\!{\beta ^2}\!{\kappa ^2}\!{\gamma _{th}}}}{{16\psi_i{\mu _2}}}}\!.
\end{split}
\label{poutGG}
\end{equation}

When the pointing errors are neglected, i.e., $ \xi \to \infty $, the FSO hop is only affected by GG atmospheric turbulence. After using \cite[(07.34.25.0007.01),
(07.34.25.0006.01) and (06.05.16.0002.01)]{sajt} to find the limit of (\ref{poutGG}) for $ \xi \to \infty $, and assuming that the relay with the best estimated CSI is always available $( M=l )$, the result in (\ref{poutGG}) is simplified to the corresponding one in \cite[(15)]{telfor}.

If system consists of only one relay, the outage probability
is derived by substituting $M=l=1$ into (\ref{pout}) as
\begin{equation}
\begin{split}
&{P_{out}^{M=1}}\!= \!1\! -\!{e^{ - \frac{{{\gamma _{th}}}}{{ {\mu _1}}}}}\!+\!\sum\limits_{k = 1}^\beta\sum\limits_{t = 0}^\infty \sum\limits_{d = 0}^t{\frac{{{t \choose d}{\rho ^t}{\xi ^2}{2^{\alpha  + k - 4}}}}{{\pi t{!^2} \left( {1 - \rho } \right)^{-1}}}} \\
&\times {\rm A}{a_k}{\left( {\frac{{\alpha \beta }}{{g\beta  + \Omega '}}} \right)^{ - \frac{{\alpha  + k}}{2}}} \!\!\!{\left( {\frac{{{\gamma _{th}}}}{{\left( {1 - \rho } \right){\mu _1}}}} \right)^{t - d}} \!\!{e^{ - \frac{{{\gamma _{th}}}}{{\left( {1 - \rho } \right){\mu _1}}}}} \\
& \! \!\times \!\MeijerG*{6}{2}{3}{7}{1, \, -t, \, {\frac{{{\xi^2} + 2}}{2}}}{{\frac{{{\xi ^2} }}{2}},\! \, \!{\frac{{\alpha}}{2},}\! \,\!{\frac{{\alpha+1}}{2},} \!\,\!{\frac{{k }}{2},} \!\,\!{\frac{{ k+1}}{2},} \!\,\!1+d,\,0}{\frac{{{\alpha ^2}{\beta ^2}{\kappa ^2}{{\left( {g \!+ \!\Omega '} \right)}^2}{\gamma _{th}}}}{{16{{\left( {g\beta  + \Omega '} \right)}^2}{\mu _2}}}}\!.
\end{split}
\label{poutM1}
\end{equation}

When system consists of only one relay and the second FSO hop is affected by  the GG atmospheric turbulence with the pointing errors, the outage probability is derived by substituting $M=l=1$ into  (\ref{poutGG}), which represents the result already reported in \cite[(23)]{JSAC2}.

\subsection{High SNR Approximations}

When the value of the electrical SNR of the FSO link is very high, the outage probability for any value of $\mu_1$ can be derived by taking the limit of (\ref{pout}), i.e.,  $ P_{out}^{{\mu _2}}=\mathop {\lim }\limits_{{\mu _2} \to \infty } {P_{out}}$. Observe that the Meijer's $ G $-function is the only term dependent on $ {{\mu _2} } $ in (\ref{pout}). After utilizing  \cite[(07.34.06.0001.01)]{sajt}, it can be concluded that the Meijer's $ G $-function tends to zero when $ {\mu _2}\! \to \!\infty $. The high electrical SNR approximation is derived as
\begin{equation}
P_{out}^{{\mu _2}}\!\! =\!\!\!\! \mathop {\lim }\limits_{{\mu _2} \to \infty } \!\!\!\!{P_{out}}\!\!\approx\!\!1\! -\! l {M \choose l}\!\!\sum\limits_{i = 0}^{l - 1}   \frac{{{{{l-1 \choose i}\left( { - 1} \right)}^i}{e^{ - \frac{{{\gamma _{th}}\left( {M\!-\!l\!+\!i\!+\!1} \right)}}{{\psi_i {\mu _1}}}}}}}{{M - l + i + 1}}\!.
\label{mi2inf}
\end{equation}

When $\mu_1$ tends to infinity, the third addend in (\ref{pout}) is nonzero only when $ t=d $. Considering this, as well as $\mathop {\lim }\limits_{{\mu _1} \to \infty }{e^{ - \frac{{{\gamma _{th}}\left( {M - l + i + 1} \right)}}{{\psi_i {\mu _1}}}}}=1  $ and $\mathop {\lim }\limits_{{\mu _1} \to \infty }{e^{ - \frac{{{\gamma _{th}}}}{{\left( {1 - \rho } \right){\mu _1}}}}}=1$, the high average SNR approximation is obtained as
\begin{equation}
\begin{split}
&P_{out}^{{\mu _1}}\!\! =\!\!\mathop {\lim }\limits_{{\mu _1} \to \infty } \!\!\!\!{P_{out}}\approx1\!-\!l {M \choose l}\!\!\sum\limits_{i = 0}^{l - 1}  \frac{{{{{l-1 \choose i} \left( { - 1} \right)}^i}}}{{M - l + i + 1}} \\
& \!+ l{M \choose l}\!\!\sum\limits_{k = 1}^\beta  {\sum\limits_{i = 0}^{l - 1} {\sum\limits_{t = 0}^\infty {\frac{{{{ {l-1 \choose i}\left( { - 1} \right)}^i}{\rho ^t}(1-\rho){\psi ^{ - (t + 1)}}}}{{\pi t{!^2}}}}} } \\
&\! \times {\rm A}{a_k}{\left( {\frac{{\alpha \beta }}{{g\beta  + \Omega '}}} \right)^{ - \frac{{\alpha  + k}}{2}}}{\xi ^2}{2^{\alpha  + k - 4}}\\
& \!\times \!\MeijerG*{6}{2}{3}{7}{1, \, -t, \, {\frac{{{\xi ^2} + 2}}{2}}}{{\frac{{{\xi ^2} }}{2}},\! \, \!{\frac{{\alpha}}{2},} \!\,\!{\frac{{\alpha+1}}{2},}\! \,\!{\frac{{k }}{2},}\! \,\!{\frac{{ k+1}}{2},} \!\,1+t,\,0}{\frac{{{\alpha ^2}{\beta ^2}{\kappa ^2}{{\left( {g + \Omega '} \right)}^2}{\gamma _{th}}}}{{16{{\left( {g\beta  + \Omega '} \right)}^2}\psi_i{\mu _2}}}}\!.
\end{split}
\label{mi1inf}
\end{equation}
Based on (\ref{mi2inf}) and (\ref{mi1inf}), the outage probability floors can be efficiently calculated. The meaning of outage floor is that the further increase of the signal power will not improve system performance, which will be illustrated in the next Section.

Since the approximation in (\ref{mi1inf}) is represented in the infinite series form, the simpler outage probability approximation when  $\mu_1$ tends to infinity is obtained by considering only the first term of infinite summation in (\ref{mi1inf}) as
\begin{equation}
\begin{split}
&P_{out}^{{\mu _{1_{app}}}}\!\!\approx1\!-\!l {M \choose l}\!\!\sum\limits_{i = 0}^{l - 1}   \frac{{{{{l-1 \choose i}\left( { - 1} \right)}^i}}}{{M - l + i + 1}} \\
& \!\! \!\!+ l{M \choose l}\!\!\sum\limits_{k = 1}^\beta \! {\sum\limits_{i = 0}^{l - 1} {\frac{{{{{l-1 \choose i}\left( { - 1} \right)}^i}{\xi ^2}{2^{\alpha \!+ \!k\! -\! 4}}{\rm A}{a_k}}}{{\pi\psi_i (1-\rho)^{-1}}}}} {\left( {\frac{{\alpha \beta }}{{g\beta\! +\! \Omega '}}} \right)^{\!\!\!-\frac{{\alpha\!+\!k}}{2}}}\!\!\! \\
& \!\times \!\MeijerG*{6}{2}{3}{7}{1, \, 0, \, {\frac{{{\xi ^2} + 2}}{2}}}{{\frac{{{\xi ^2} }}{2}}, \, {\frac{{\alpha}}{2},} \,{\frac{{\alpha+1}}{2},} \,{\frac{{k }}{2},} \,{\frac{{ k+1}}{2},} \,1,\,0}{\frac{{{\alpha ^2}{\beta ^2}{\kappa ^2}{{\left( {g + \Omega '} \right)}^2}{\gamma _{th}}}}{{16{{\left( {g\beta  + \Omega '} \right)}^2}\psi_i{\mu _2}}}}\!.
\end{split}
\label{mi1inf1}
\end{equation}
This approximation is valid only in certain conditions. Since the infinite series in (\ref{mi1inf}) originates from the representation of the modified
Bessel function of the first kind into a series form (see Appendix \ref{App1}), the approximation in (\ref{mi1inf1}) is valid when the argument of $ I_0(.) $  tends to zero, i.e., for lower values of  $\rho $.

\section{Numerical results and simulations}

In this Section, numerical results obtained by derived  outage probability expressions  are presented. In addition, Monte Carlo simulations are provided to confirm the accuracy of the derived expressions. The second FSO hop is influenced by the $ \mathcal{M} $-distributed atmospheric turbulence channel when the pointing errors is taken into account. Based on \cite{book}, the intensity of atmospheric turbulence is determined by the Rytov variance, previously defined as $ \sigma_R^{2}=~1.23C_n^{2}\iota^{7/6}d^{11/6} $, with the  refractive index structure parameter, $ C_n^{2} $, varying in the range from $ 10^{-17} $ to $ 10^{-13} $ m$ ^{-2/3} $. In this paper,  the values of the parameters are taken from \cite{M1,M4,M5}, which are determined by some experimental measurements. The FSO link is assumed to be $ 1 $ km long, and the wavelength employed in optical second hop is $ 785 $ nm. In addition, the average optical power of the FSO hop is normalized, i.e., $\Omega+2b_0=1 (\Omega=0.5,b_0=0.25)$, and $ {{\phi _A} - {\phi _B}}=\pi/2 $. The pointing errors strength is determined by the normalized jitter standard deviation, $\sigma_s/a$, where the radius of a circular detector aperture takes a value $a=5$ cm. Further, the value of the optical beam radius at the waist is $ a_0=5 $ cm, and of the radius of curvature is $ F_0=-10$ \cite{PE4,chapter}. The value of the outage threshold is $ \gamma_{th}=-10$ dB.

\begin{table}[b]
\centering
\caption{ Values of  $\alpha,\beta, \rho_M$ for different scattering component $U_S^C $ strength for the same intensity of atmospheric turbulence  ($ \sigma_R^{2}=0.36$,  $C_n^2=0.83 \times 10^{-14} $ m$ ^{-2/3} $) \cite{M5}}
\begin{tabular}{cc}
\hline 
$ (\alpha,\beta, \rho_M) $ & impact of $ U_S^C  $   \\
\hline
$ (11,4,1) $ & low $(g=0)$  \\
$ (10,5,0.95) $ & medium  \\
$ (25,10,0.75) $ & great  \\
\hline
\end{tabular}
  \label{table}
\end{table} 

\begin{figure}[!b]
\centering
\includegraphics[width=3.5in]{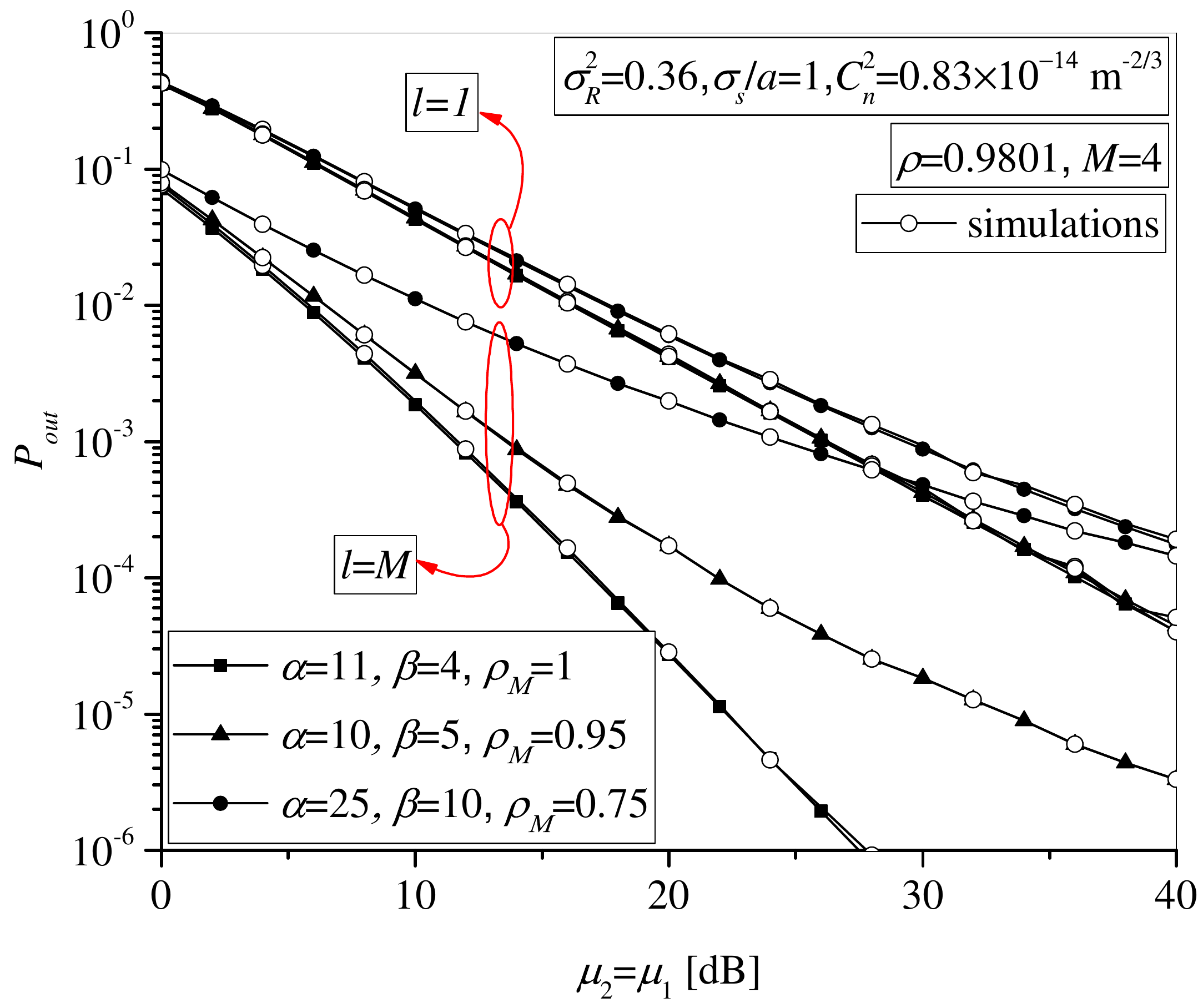}
\caption{Outage probability vs. $ \mu_1=\mu_2 $ when the best and the worst relay is selected to perform transmission.}
\label{Fig_22}
\end{figure}

Fig.~\ref{Fig_22} presents the outage probability dependence on $ \mu_1~=~\mu_2 $ when the relay with the best estimated CSI is able to perform further transmission $ (l=M) $, as well as when all except the one with worst estimated CSI are unavailable $ (l=1) $. The atmospheric turbulence intensity is determined to be $  \sigma_R^{2}=0.36 $ and $ C_n^2=0.83 \times 10^{-14} $ m$ ^{-2/3} $,  based on  experimental measurements performed in University of Waseda, Japan, on the 15th October, 2009 (see \cite{M5}). 
For the same Rytov variance, the following parameters $ (\alpha,\beta, \rho_M) $ are considered: $ (11,4,1) $, $ (10,5,0.95) $, and $ (25,10,0.75) $, considering different strength of the scattering component  $ {U_S^G} $ (see Table I). Since the intensity of the turbulence is the same, in Fig.~\ref{Fig_22} the value of the parameter $\rho_M$, representing the amount of the scattering power coupled to the LOS component, defines the outage probability performance. Greater values of $ \rho_M $ reflect in improved system performance, meaning that the average power of the scattering component $ {U_S^G} $, $ g $, is lower. In fact, when $ \rho_M =1 $, it holds that $ g=0 $, i.e., total scattering power is related to the component $ {U_S^C}  $. This case implies that the atmospheric turbulence is modeled by GG distribution, which does not take into consideration  the scattering component received by off-axis eddies. Lowering the value of $ \rho_M $, the average power $ g $ is greater and the scattered component $ {U_S^G} $ has greater impact on the system performance. Furthermore, it is expected that the outage probability is lower when the selected relay is the one with best estimated CSI.

\begin{figure}[!b]
\centering
\includegraphics[width=3.5in]{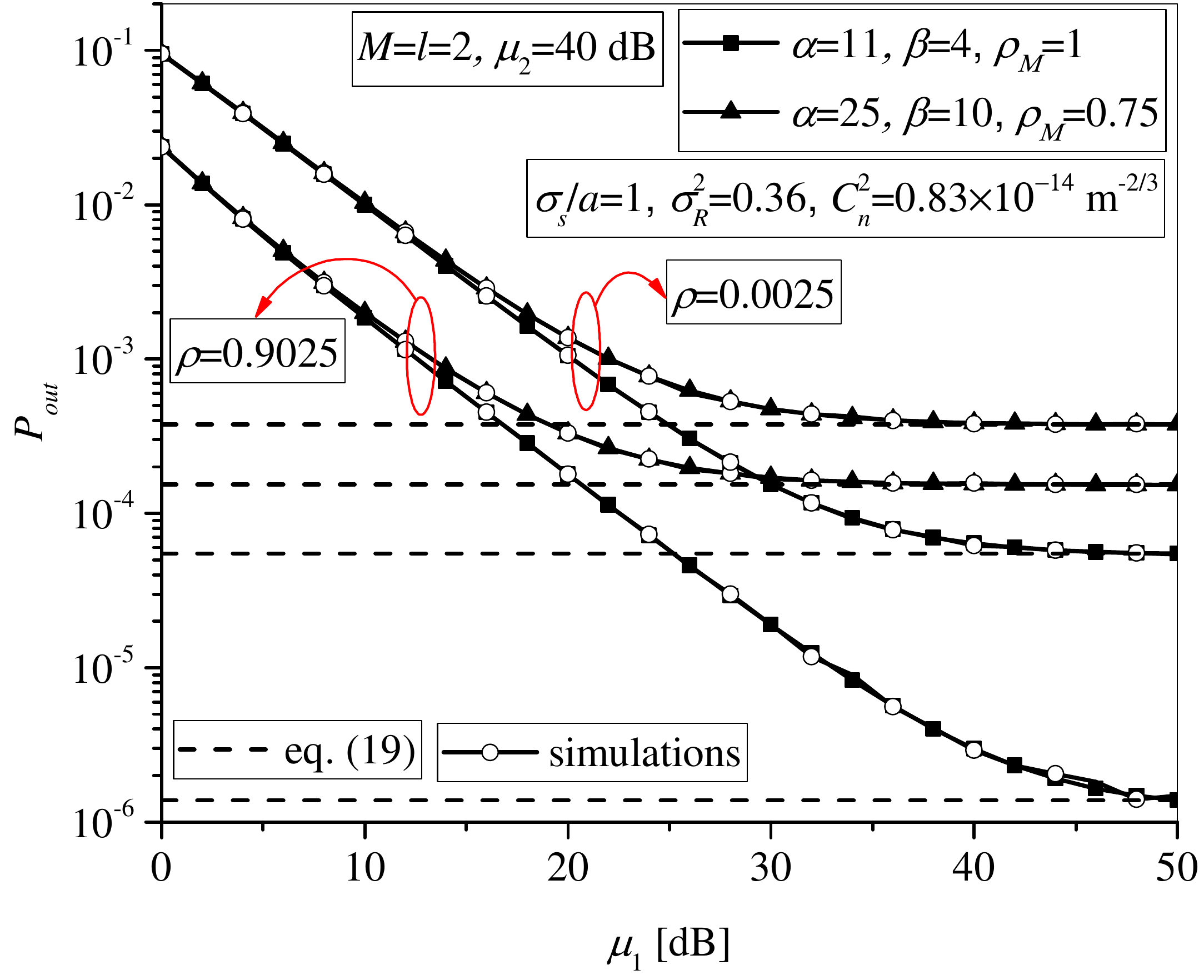}
\caption{Outage probability vs. $ \mu_1 $ for different values of correlation coefficient and the amount of the scattering power coupled to the LOS component.}
\label{Fig_21}
\end{figure}

\begin{figure}[!t]
\centering
\includegraphics[width=3.5in]{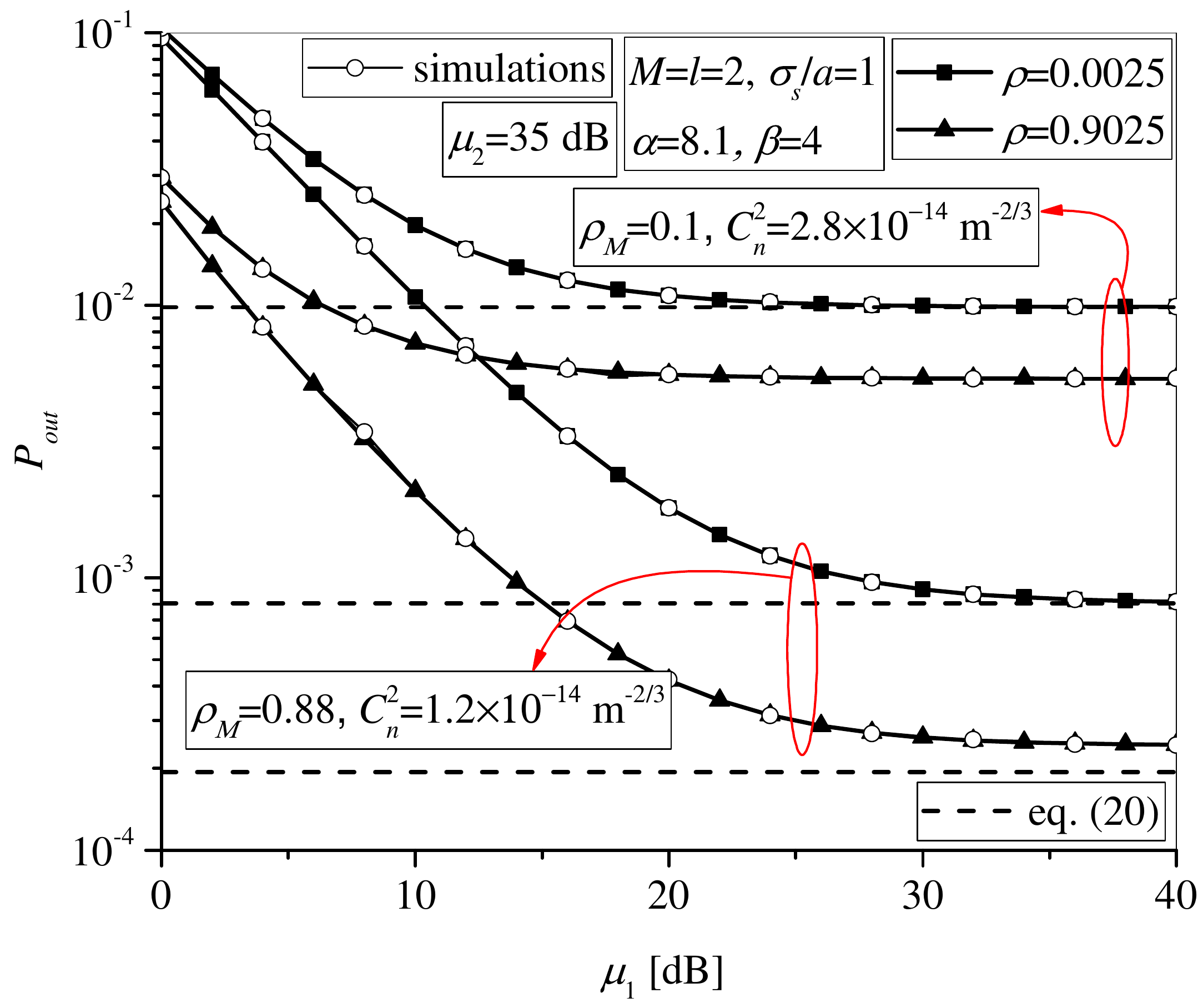}
\caption{Outage probability vs. $ \mu_1 $ for different values of correlation coefficient in various atmospheric turbulence conditions.}
\label{Fig_24}
\end{figure}

\begin{figure}[!t]
\centering
\includegraphics[width=3.5in]{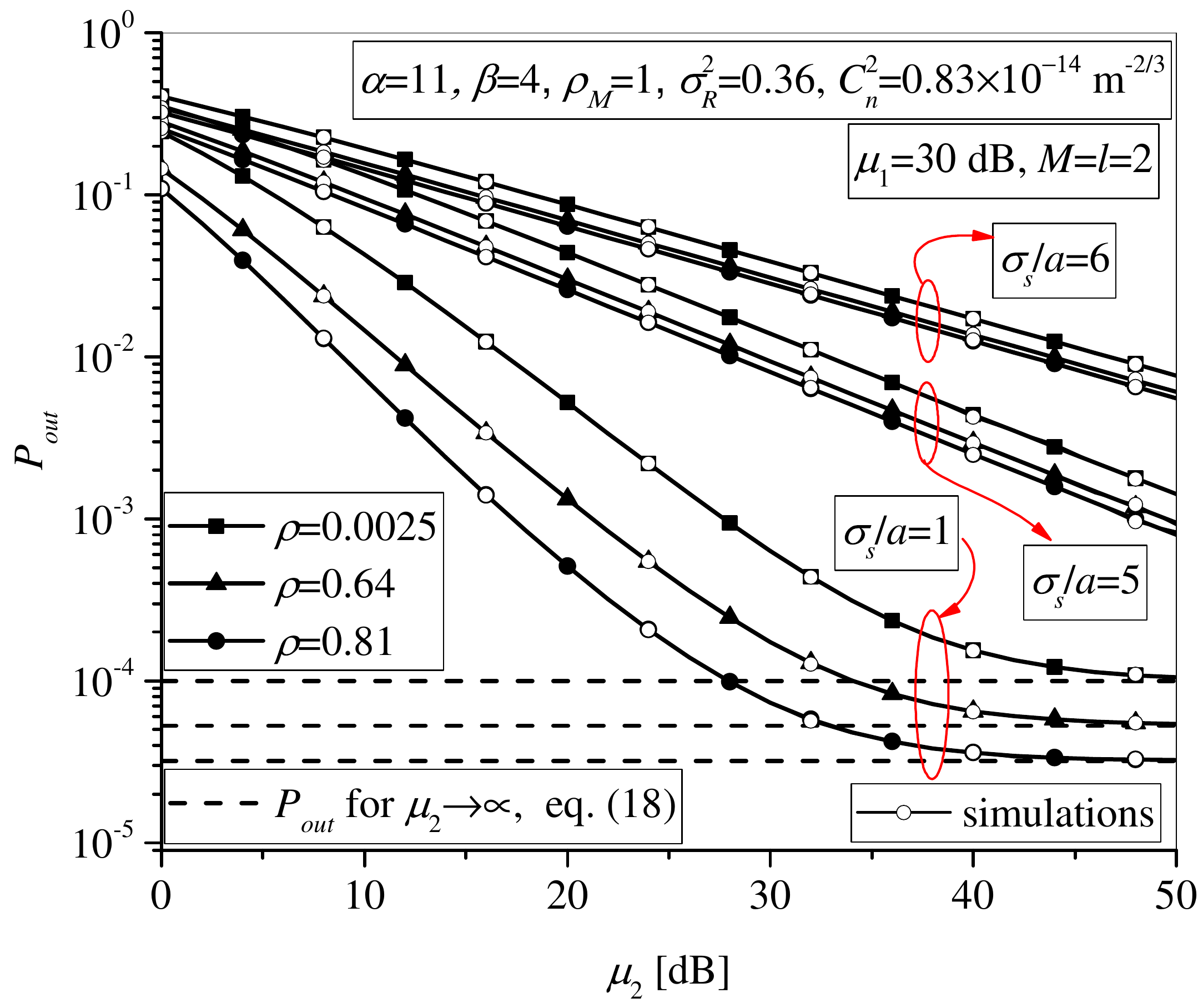}
\caption{Outage probability vs. $ \mu_2 $ for different values of correlation coefficient and various pointing errors strength.}
\label{Fig_17b}
\end{figure}

The outage probability dependence on the electrical SNR of the FSO hop for the same intensity of the atmospheric turbulence is presented in Fig.~\ref{Fig_21}. Different values of the amount of the scattering power coupled to the LOS component, $ \rho_M $, and the correlation coefficient, $ \rho $, are assumed. Greater values of the parameter $\rho$ mean that the outdated CSI, used for PRS and relay gain determination, and actual CSI of RF link at the time of transmission, are more correlated, leading to the better system performance. The impact of $ \rho_M $ on the outage probability is more expressed when the value of $\rho$ is greater. When the correlation coefficient is lower, the RF channel conditions are worse,  and the impact  of the scattering conditions of the FSO hop on system performance is lower. 

In addition, the results obtained based on the high SNR approximation in (\ref{mi1inf}), are also presented in Fig~\ref{Fig_21}. When the average SNR over RF link is very great, the outage probability floor exists, and further increase in signal power at the source node  will not result in improved system performance. The results obtained by (\ref{mi1inf}) are overlapped with the results achieved by (\ref{pout}) for greater $ \mu_1 $. Further, the outage probability floor occurs at lower vales of $ \mu_1 $ when $\rho $ or  $\rho_M$ is lower.

\begin{table}[b]
\centering
\caption{ Values of  $\sigma_R,C_n^2, \rho_M$ for different intensity of the  atmospheric turbulence  ($\alpha=8.1$,  $\beta=4 $ ) \cite{M4}}
\begin{tabular}{cc}
\hline 
$ (\sigma_R,C_n^2, \rho_M) $ &  atmospheric turbulence  strength \\
\hline
$ (0.52, 1.2 \times 10^{-14}$ m$ ^{-2/3} $, 0.88) & weak  (in sunrise)  \\
$ (1.2, 2.8 \times 10^{-14}$ m$ ^{-2/3} $, 0.1)  & strong (in mid-day)  \\
\hline
\end{tabular}
  \label{table}
\end{table}

Fig.~\ref{Fig_24} shows the outage probability dependence on $ \mu_1$ for different values of parameter  $\rho $. The atmospheric turbulence is determined by the Rytov variance and the refractive index structure parameter  given in Table II. 
As expected, system performs better in weak turbulence conditions. Further, the correlation effect on outage probability is more pronounced when the second FSO hop is influenced by weak atmospheric turbulence.

The  high SNR approximation results obtained based on (\ref{mi1inf1}) are also provided in Fig.~\ref{Fig_24}. The outage probability floor is  present for great values of $ \mu_1$, which is in agreement with the curves obtained based on (\ref{mi1inf1}), but only when the correlation coefficient is low. 

The impact of the pointing errors strength on the outage probability is depicted in Fig.~\ref{Fig_17b}, assuming different values of parameter  $\rho $. System has better performance when the normalized jitter standard deviation is lower, meaning that the alignment between transmitter laser at the relay and receiver telescope at the destination is very good and the pointing error is low. The impact of the correlation on the overall outage probability is more prominent when the pointing error is low.

In addition, the outage probability floor exists for great values of the electrical SNR over FSO link. Further increase of the optical  power will not improve system performance. This outage probability floor  is in agreement with the high electrical SNR approximation results  obtained, based on (\ref{mi2inf}). As it can be concluded from both (\ref{mi2inf}) and Fig.~\ref{Fig_17b}, this outage probability floor is not dependent on the FSO channel conditions, but only on RF channel parameters. With increasing the electrical SNR over FSO link, the curves for $\sigma_s/a=1 $, $\sigma_s/a=5 $ and $\sigma_s/a=6 $ when the correlation coefficient is the same, will overlap.

\begin{figure}[!t]
\centering
\includegraphics[width=3.5in]{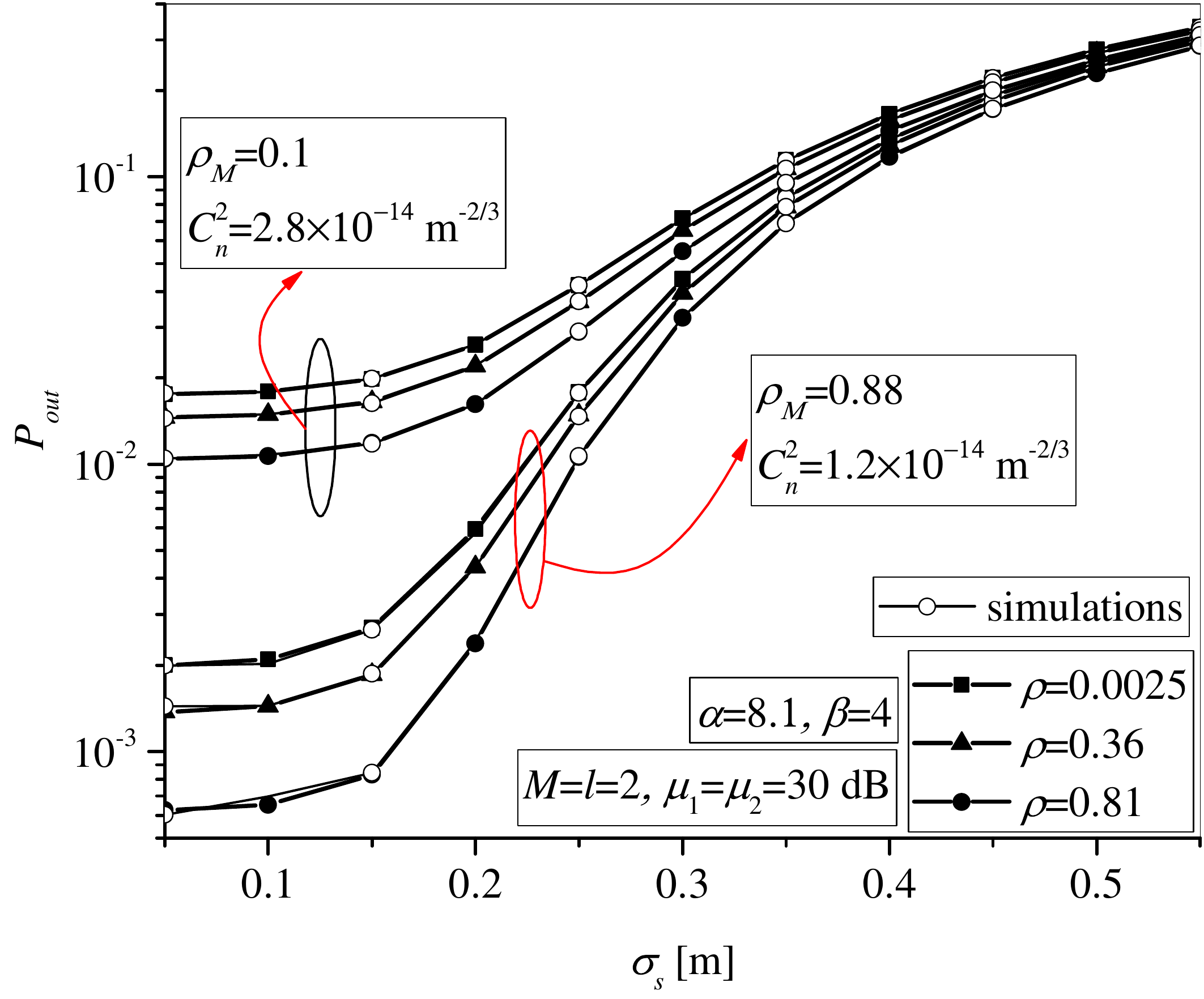}
\caption{Outage probability vs. $ \sigma_s $ for different values of correlation coefficient.}
\label{Fig_11}
\end{figure}

\begin{figure}[!t]
\centering
\includegraphics[width=3.5in]{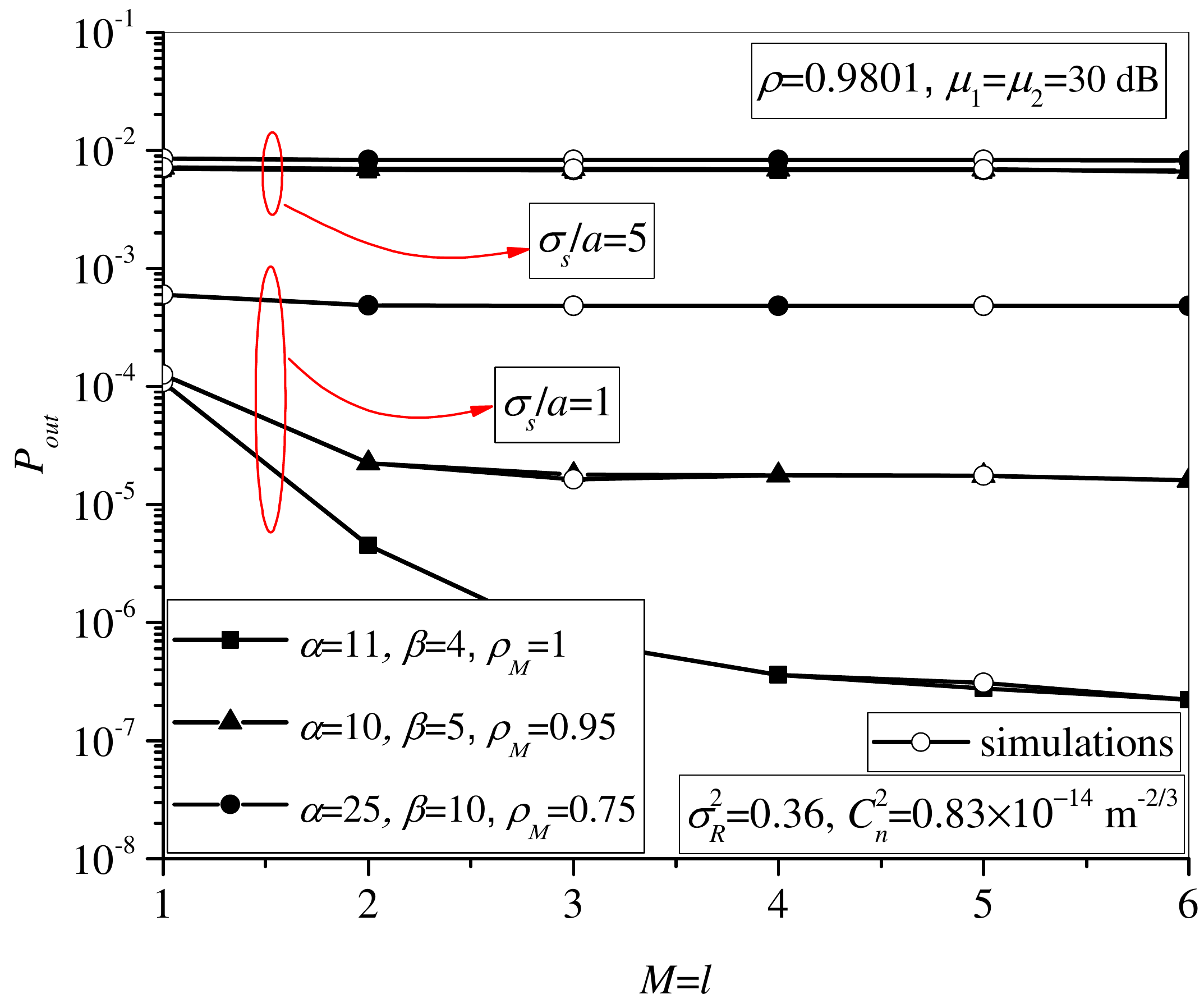}
\caption{Outage probability vs. number of relays  $ M$.}
\label{Fig_19}
\end{figure}

Fig.~\ref{Fig_11} presents the outage probability dependence on jitter standard deviation. In accordance with the conclusions from Fig.~\ref{Fig_24} and Fig.~\ref{Fig_17b}, the effect of correlation has more influence on the outage probability in weak atmospheric turbulence condition and when  the pointing error is low. In addition, the pointing errors effect is more dominant when the optical signal transmission suffers from weak atmospheric turbulence.

The usefulness of the multiple relay implementation within RF/FSO system in regards to various atmospheric turbulence and pointing errors conditions is presented in Fig.~\ref{Fig_19}. The greatest SNR gain is accomplished by using PRS with two relays in regard to one relay. Also, when the FSO hop suffers from damaging conditions (strong atmospheric turbulence or intense pointing errors),  the multiple relays implementation within RF/FSO system will not provide significant system performance improvement.

\section{Conclusion}
This paper has presented the outage probability analysis of the mixed multiple AF relying RF/FSO system. Contrary to previous published studies, this system has employed the variable gain AF relays, and the selection of the active relay is performed based on PRS procedure. The RF links experience the Rayleigh fading environment, which is characterized by fast fading statistics. For that reason, the outdated CSI of the RF channel is used for PRS and relay gain determination. The intensity fluctuations of the optical signal are assumed to originate from M$\acute{{\rm{a}}}$laga ($ \mathcal{M} $) distribution, as well as from pointing errors. The outage probability expression is derived, and simplified to some special cases previously reported in literature. Approximate high SNR expressions are also provided. 

Based on derived expressions, numerical results have been presented and confirmed by Monte Carlo simulations. The effects of system and channel parameters have been examined. The existence of the outage probability floor has been observed, which is an important limiting factor of the RF/FSO system. The outage floor can be efficiently calculated by derived approximate expressions in high average/electrical SNR region. It has  been illustrated that the outdated CSI used for relay gain adjustment and PRS procedure can seriously determine the outage probability performance, particularly when the optical signal transmission via FSO hop is endangered by favorable conditions (weak atmospheric turbulence in the mid-day, very low average power of the scattering component received by off-axis eddies, and/or weak pointing errors). Furthermore, the pointing errors phenomenon is more important to the system performance, when transmission is performed via weak atmospheric turbulence conditions.

As the most significant conclusion obtained by presented analysis and results, we have concluded that the implementation of multiple relays within RF/FSO system will not provide important performance improvement compared with the costs and difficulty of implementation 
 when the optical signal transmission via free space in the second hop is impaired by harmful conditions, such as strong atmospheric turbulence or expressive misalignment between FSO apertures.

\section*{Acknowledgement}
This work has received funding from the European Union Horizon 2020 research and innovation programme under the Marie Skodowska-Curie grant agreement No 734331. The work was  supported by Ministry of Education, Science and Technology Development of Republic of Serbia under grants TR-32025 and III-47020.

\appendices
\section{}
\label{App1}
After substituting (\ref{pdf_rf}) and (\ref{cpdf_g2}) into (\ref{pout2}), the outage probability is expressed as
\begin{equation}
{P_{out}} = 1 - {\Im _1} + {\Im _2},
\label{A1}
\end{equation}
where integral $ {\Im _1} $ is defined and solved by \cite[(6.614.3)]{Grad}, \cite[(07.44.03.0007.01) and (07.02.03.0002.01)]{sajt} as
\begin{equation}
\begin{split}
&{\Im _1} = l{M \choose l}\sum\limits_{i = 0}^{l - 1} {l-1 \choose i}\frac{{{{\left( { - 1} \right)}^i}}}{{\left( {1 - \rho } \right)\mu _1^2}}{e^{ - \frac{{{\gamma _{th}}}}{{\left( {1 - \rho } \right){\mu _1}}}}}\\
& \times \!\!\!\int\limits_0^\infty \!\!\! {\int\limits_0^\infty  {{e^{ - \frac{x}{{\left( {1 - \rho } \right){\mu _1}}}}}{e^{ - \frac{{\psi_ y}}{{\left( {1 - \rho } \right){\mu _1}}}}}} {I_0}}\!\! \left( {\frac{{2\sqrt {\rho \left( {x + {\gamma _{th}}} \right)y} }}{{\left( {1 - \rho } \right){\mu _1}}}} \right)\!\!dxdy\\
& = \! l{M \choose l}\!\sum\limits_{i = 0}^{l - 1} {l-1 \choose i} \frac{{{{\left( { - 1} \right)}^i}}}{{\left( {M - l + i + 1} \right)}}{e^{ - \frac{{{\gamma _{th}}\left( {M - l + i + 1} \right)}}{{\psi_i {\mu _1}}}}}.
\end{split}
\label{A2}
\end{equation}
Integral $ {\Im _2} $ is defined as
\begin{equation}
\begin{split}
&{\Im _2}\!= l{M \choose l}\!\!\sum\limits_{k = 1}^\beta  {\sum\limits_{i = 0}^{l - 1} \frac{{{{{l-1 \choose i} \left( { - 1} \right)}^i}{\xi ^2}{\rm A}{a_k}}}{{2\left( {1 - \rho } \right)\mu _1^2}}} {\left(\! {\frac{{\alpha \beta }}{{g\beta  + \Omega '}}}\! \right)^{\!\! \!\!-\! \frac{{\alpha  + k}}{2}}}{\kern 1pt} \\
&\!\! \times \!\!{\kern 1pt} {e^{ - \frac{{{\gamma _{th}}}}{{\left( {1 - \rho } \right){\mu _1}}}}}\!\!\!\!\int\limits_0^\infty\!\! \!\! {\int\limits_0^\infty \!\! {{e^{ - \frac{x}{{\left( {1 - \rho } \right){\mu _1}}}}}} } {e^{ - \frac{{\psi_i y}}{{\left( {1 - \rho } \right){\mu _1}}}}}{I_0}\!\!\left( {\frac{{2\sqrt {\rho \left( {x\!+ \!{\gamma _{th}}} \right)y} }}{{\left( {1 - \rho } \right){\mu _1}}}} \right)\\
&\times \MeijerG*{3}{1}{2}{4}{1, \, \xi^2+1}{\xi ^2, \, \alpha, \, k, \, 0}{\frac{{\alpha \beta \kappa \left( {g + \Omega '} \right)}}{{g\beta  + \Omega '}}{\sqrt {\frac{{{\gamma _{th}}y}}{{{\mu _2}x}}} }} dxdy. 
\end{split}
\label{A3}
\end{equation}
After utilization of \cite[(03.02.06.0037.01)]{sajt} as $ {I_0}\! \! \left( {\frac{{2\sqrt {\rho \left( {x + {\gamma _{th}}} \right)y} }}{{\left( {1 - \rho } \right){\mu _1}}}} \right)\!\! \! \!  = \! \! \! \!  \sum\limits_{t = 0}^\infty  {\frac{{{\rho ^t}{{\left( {x + {\gamma _{th}}} \right)}^t}{y^t}}}{{t{!^2}{{\left( {1 - \rho } \right)}^{2t}}\mu _1^{2t}}}}  $, integral $ \! {\Im _2} $ is re-written as
\begin{equation}
\begin{split}
&{\Im _2}\! =\! l{M \choose l}\!\sum\limits_{k = 1}^\beta  {\sum\limits_{i = 0}^{l - 1}  \frac{{{{{l-1 \choose i}\left( { - 1} \right)}^i}{\xi ^2}{\rm A}{a_k}}}{{2\left( {1 - \rho } \right)\mu _1^2}}} {\left(\! {\frac{{\alpha \beta }}{{g\beta  + \Omega '}}}\! \right)^{ \!\!\!\!- \frac{{\alpha  + k}}{2}}}{\kern 1pt} \\
& \!\!\times \!\!\sum\limits_{t = 0}^\infty \! {\frac{{{\rho ^t}{e^{ - \frac{{{\gamma _{th}}}}{{\left( {1 - \rho } \right){\mu _1}}}}}}}{{t{!^2}{{\left( {1 - \rho } \right)}^{2t}}\mu _1^{2t}}}} \!\!\int\limits_0^\infty \!\! {{{\left( {x + {\gamma _{th}}} \right)}^t}{e^{ - \frac{x}{{\left( {1 - \rho } \right){\mu _1}}}}}} dx \times {\Im _{21}}, 
\end{split}
\label{A4}
\end{equation}
where 
\begin{equation}
\begin{split}
{\Im _{21}}& = \int\limits_0^\infty  {{y^t}} {e^{ - \frac{{\psi_i y}}{{\left( {1 - \rho } \right){\mu _1}}}}} \\
& \times \MeijerG*{3}{1}{2}{4}{1, \, \xi^2+1}{\xi ^2, \, \alpha, \, k, \, 0}{\frac{{\alpha \beta \kappa \left( {g + \Omega '} \right)}}{{g\beta  + \Omega '}}{\sqrt {\frac{{{\gamma _{th}}y}}{{{\mu _2}x}}} }} dy. 
\end{split}
\label{A5}
\end{equation}
After representing exponential function in terms of Meijer's $ G $-function as $ {e^{ - \frac{{\psi_i y}}{{\left( {1 - \rho } \right){\mu _1}}}}} = \MeijerG*{1}{0}{0}{1}{-}{0}{{\frac{{\psi_i y}}{{\left( {1 - \rho } \right){\mu _1}}}}} $ by using \cite[(01.03.26.0004.01)]{sajt}, integral $ {\Im _{21}} $ is solved with a help of \cite[(07.34.21.0013.01)]{sajt}. The Meijer's $ G $-function in obtained expression is simplified and transformed by \cite[(07.34.03.0002.01), (07.34.03.0001.01) and (07.34.16.0002.01)]{sajt}, and the final integral $ {\Im _{21}} $ is 
\begin{equation}
\begin{split}
{\Im _{21}}&= \frac{{{2^{\alpha  + k - 3}}}}{\pi }{\left( {\frac{\psi_i }{{\left( {1 - \rho } \right){\mu _1}}}} \right)^{ - (t + 1)}} \\
& \times \MeijerG*{2}{5}{6}{3}{{\frac{2-\xi^2}{2}},\,{\frac{2-\alpha}{2}},\,{\frac{1-\alpha}{2}},\,{\frac{2-k}{2}}, \, {\frac{1-k}{2}},\,1}{0, \, 1+t, \, -\frac{\xi^2}{2}} {{\Psi x}}, 
\end{split}
\label{A6}
\end{equation}
where $ \Psi  = \frac{{16{{\left( {g\beta  + \Omega '} \right)}^2}\psi_i {\mu _2}}}{{{\alpha ^2}{\beta ^2}{\kappa ^2}{{\left( {g + \Omega '} \right)}^2}{\gamma _{th}}\left( {1 - \rho } \right){\mu _1}}} $.
Next, after substituting (\ref{A6}) into (\ref{A4}), integral $\Im _{2}$ is found as
\begin{equation}
\begin{split}
&{\Im _2} = l{M \choose l}\sum\limits_{k = 1}^\beta  {\sum\limits_{i = 0}^{l - 1} \frac{{{{ {l-1 \choose i}\left( { - 1} \right)}^i}{\xi ^2}A{a_k}{2^{\alpha  + k - 3}}}}{{2\pi }}} \\
& \!\!\times \!\!{e^{ - \frac{{{\gamma _{th}}}}{{\left( {1 - \rho } \right){\mu _1}}}}}{\left( {\frac{{\alpha \beta }}{{g\beta  + \Omega '}}} \right)^{\!\!\!\!\!- \frac{{\alpha  + k}}{2}}}\!\!\! \sum\limits_{t = 0}^\infty  {\frac{{{\rho ^t}{\psi_i ^{ - (t + 1)}}}}{{t{!^2}{{\left( {1 - \rho } \right)}^t}\mu _1^{t + 1}}}}  \times {\Im _{22}},
\end{split}
\label{A7}
\end{equation}
where 
\begin{equation}
\begin{split}
{\Im _{22}}&= \int\limits_0^\infty  {{{\left( {x + {\gamma _{th}}} \right)}^t}{e^{ - \frac{x}{{\left( {1 - \rho } \right){\mu _1}}}}}} \\
& \times \MeijerG*{2}{5}{6}{3}{{\frac{2-\xi^2}{2}},\,{\frac{2-\alpha}{2}},\,{\frac{1-\alpha}{2}},\,{\frac{2-k}{2}}, \, {\frac{1-k}{2}},\,1}{0, \, 1+t, \, -\frac{\xi^2}{2}} {{\Psi x}}.  
\end{split}
\label{A8}
\end{equation}
Binomial theorem \cite[(1.111)]{Grad} is applied as $ {\left( {x + {\gamma _{th}}} \right)^t}\! \! \!  = \sum\limits_{d = 0}^t {t \choose d} {x^d}\gamma _{th}^{t - d} $, and by using \cite[(01.03.26.0004.01)]{sajt} the exponential function is represented in terms of Meijer's $ G $-function as $ {e^{ - \frac{{x}}{{\left( {1 - \rho } \right){\mu _1}}}}}\! \! =\! \!  \MeijerG*{1}{0}{0}{1}{-}{0}{{\frac{{x}}{{\left( {1 - \rho } \right){\mu _1}}}}} $. Afterwards, integral $\!  {\Im _{22}} $ is solved with the help of \cite[(07.34.21.0011.01)]{sajt} 
\begin{equation}
\begin{split}
&{\Im _{22}} = \sum\limits_{d = 0}^t {t \choose d} \gamma _{th}^{t - d}{\left( {1 - \rho } \right)^{d + 1}}{\mu _1}^{d + 1} \\
&\times \!\MeijerG*{2}{6}{7}{3}{{\frac{2-\xi^2}{2}},\,{\frac{2-\alpha}{2}},\,{\frac{1-\alpha}{2}},\,{\frac{2-k}{2}}, \, {\frac{1-k}{2}},\,-d,\,1}{0, \, 1+t, \, -\frac{\xi^2}{2}} {{\Psi \left( {1 - \rho } \right){\mu _1}\!}}.   
\end{split}
\label{A9}
\end{equation}
The  Meijer's $ G $-function in (\ref{A9}) is transformed by using \cite[(07.34.16.0002.01)]{sajt} as
\begin{equation}
\begin{split}
&\MeijerG*{2}{6}{7}{3}{{\frac{2-\xi^2}{2}},\,{\frac{2-\alpha}{2}},\,{\frac{1-\alpha}{2}},\,{\frac{2-k}{2}}, \, {\frac{1-k}{2}},\,-d,\,1}{0, \, 1+t, \, -\frac{\xi^2}{2}} {\Psi \left( {1 - \rho } \right){\mu _1}}\\
& = \MeijerG*{6}{2}{3}{7}{1, \, -t, \, \frac{\xi^2+2}{2}}{{\frac{\xi^2}{2}},\,{\frac{\alpha}{2}},\,{\frac{\alpha+1}{2}},\,{\frac{k}{2}}, \, {\frac{k+1}{2}},\,1+d,\,0} {\frac{1}{{\Psi \left( {1 - \rho } \right){\mu _1}}}}.   
\end{split}
\label{A10}
\end{equation}

First, substitutions of (\ref{A10}) into (\ref{A9}), and afterwards (\ref{A9})  into (\ref{A7}) are performed. Next, replacement of $\Psi$ is done. Novel obtained form of $ {\Im _{2}}  $,  together with $ {\Im _{1}}  $ in (\ref{A2}), are placed in (\ref{A1}). Final outage probability expression  is presented in (\ref{pout}).

\ifCLASSOPTIONcaptionsoff
  \newpage
\fi


\begin{thebibliography}{1}
\IEEEtriggeratref{26}

\bibitem{book}
Z. Ghassemlooy, W. Popoola, and S. Rajbhandari, \emph{Optical Wireless Communications: System and Channel Modelling with MATLAB\textregistered}. Boca Raton, USA: CRC Press, 2013.

\bibitem{survey}
M. A. Khalighi and M. Uysal, "Survey on free space optical communication: A communication theory perspective," \emph{IEEE Commun. Surveys Tuts.,} vol. 16, no. 4, pp. 2231--2258, Fourthquarter 2014.

\bibitem{new1}
G. C. Mandal, R. Mukherjee, B. Das, and A. S. Patra, "A full-duplex WDM hybrid fiber-wired/fiber-wireless/fiber-VLC/fiber-IVLC transmission system based on a self-injection locked quantum dash laser and a RSOA," \emph{Opt. Commun.,}  vol. 427, pp. 202--208, Nov. 2018.

\bibitem{new2}
G. C. Mandal, R. Mukherjee, B. Das, and A. S. Patra, "Next-generation bidirectional Triple-play services using RSOA based WDM Radio on Free-Space Optics PON," \emph{Opt. Commun.,} vol. 411, pp. 138--142, Mar. 2018.

\bibitem{PE2}
A.~A.~Farid and S.~Hranilovic, "Outage capacity optimization for free space optical links with pointing errors," \emph{J. Lightw. Technol.}, vol. 25, no. 7, pp. 1702--1710, Jul. 2007.

\bibitem{PE4}
A.~A.~Farid and S.~Hranilovic, "Outage capacity for MISO intensity-modulated free-space optical links with misalignment," \emph{IEEE/OSA J. Opt. Commun. Netw.}, vol. 3, no. 10, pp.~780--789, Oct. 2011.

\bibitem{lee}
E. Lee, J. Park, D. Han, and G. Yoon, "Performance analysis of the asymmetric dual-hop relay transmission with mixed RF/FSO links," \emph{IEEE Photon. Technol. Lett.}, vol. 23, no. 21, pp. 1642--1644, November 2011.

\bibitem{endend}
H. Samimi and M. Uysal, "End-to-end performance of mixed RF/FSO transmission systems," \emph{IEEE/OSA J. Opt. Commun. Netw.}, vol. 5, no. 11, pp. 1139--1144, Nov. 2013. 

\bibitem{Ansari-Impact}
I. S. Ansari, F. Yilmaz, and M.$ - $S. Alouini, "Impact of pointing errors on the performance of mixed RF/FSO dual-hop transmission systems," \emph{IEEE Wireless Commun. Lett.}, vol. 2, no. 3, pp. 351--354, Jun. 2013.

\bibitem{Anees2}
S. Anees and M. R. Bhatnagar, "Performance of an amplify-and-forward dual-hop asymmetric RF/FSO communication system," \emph{IEEE/OSA J. Opt. Commun. Netw.}, vol. 7, no. 2, pp. 124--135, Jun. 2015.

\bibitem{Zhang_JLT}
Z. Jiayi, D. Linglong, Z. Yu, and W. Zhaocheng, "Unified performance analysis of mixed radio frequency/free-space optical dual hop transmission systems," \emph{J. Lightw. Technol.}, vol. 33, no. 11, pp. 2286--2293,  2015.


\bibitem{Zedini_PhotonJ}
E. Zedini, I. S. Ansari, and M.--S. Alouini, "Performance analysis of mixed Nakagami-m and Gamma-Gamma dual-hop FSO transmission systems," \emph{IEEE Photon. J.}, vol. 7, no. 1, pp. 1--20, Feb. 2015.

\bibitem{nova}
H. Khanna, M. Aggarwal, and S. Ahuja, "Outage analysis of a variable-gain amplify and forward relayed mixed RF-FSO system," in \emph{Proc. IEEE INDICON 2016}, Bangalore, India, 2016, pp. 1--6.

\bibitem{JSAC1}
Y. Liang, M. O. Hasna, and X. Gao, "Performance of mixed RF/FSO with variable gain over generalized atmospheric turbulence channels," \emph{IEEE J. Sel. Areas Commun.}, vol. 33, no. 9, pp. 1913--1924, Sep. 2015.

\bibitem{JSAC2}
G. T. Djordjevic, M. I. Petkovic, A. M. Cvetkovic, and G. K. Karagiannidis, "Mixed RF/FSO relaying with outdated channel state information," \emph{IEEE J. Sel. Areas Commun.}, vol. 33, no. 9, pp. 1935--1948, Sep. 2015.


\bibitem{var2}
L. Han, H. Jiang, Y. You, Z. Ghassemlooy, "On the performance of a mixed RF/MIMO FSO variable gain dual-hop transmission system," \emph{Optics Commun.}, vol. 420, pp. 59--64, June 2018.


\bibitem{var3}
H. Khanna, M. Aggarwal, and S. Ahuja, "Performance analysis of a variable-gain amplify-and-forward relayed mixed RF-FSO system," \emph{Int. J. Commun. Syst.}, vol. 31, no.1,  pp. 59--64, Aug. 20178.



\bibitem{FSO1}
M. Safari and M. Uysal, "Relay-assisted free-space optical communication," \emph{IEEE Trans. Wireless Commun.}, vol. 7, no. 12, pp. 5441--5449, Dec. 2008.



\bibitem{FSO4}
N. D. Chatzidiamantis, D. S. Michalopoulos, E. E. Kriezis, G. K. Karagiannidis, and R. Schober, "Relay selection protocols for relay-assisted free-space optical systems," \emph{IEEE/OSA J. Opt. Commun. Netw.}, vol. 5, no. 1, pp. 92--103, Jan. 2013.

\bibitem{FSO5}
J. Y. Wang, J. B. Wang, M. Chen, Y. Tang, and Y. Zhang, "Outage analysis for relay-aided free-space optical communications over turbulence channels with nonzero boresight pointing Errors," \emph{IEEE Photon. J.}, vol. 6, no. 4, pp. 1--15, Aug. 2014.

\bibitem{JLT}
M. I. Petkovic, A. M. Cvetkovic,  G. T. Djordjevic,  and G. K. Karagiannidis, "Partial relay selection with outdated channel state estimation in mixed RF/FSO systems," \emph{J. Lightw. Technol.}, vol. 33, no. 13, pp. 2860--2867, Jul. 2015.

\bibitem{PRS}
I. Krikidis, J. Thompson, S. Mclaughlin, and N. Goertz, "Amplify-and-forward with partial relay selection," \emph{IEEE Commun. Lett.}, vol. 12, no. 4, pp. 235-237, Apr. 2008.

\bibitem{chapter}
M. I. Petkovic, A. M. Cvetkovic,  and G. T. Djordjevic, "Mixed RF/FSO relaying systems," in \emph{Optical Wireless Communications -- An Emerging Technology}, Springer, 2016.

\bibitem{prsF1}
E. Balti, M. Guizani and B. Hamdaoui, "Hybrid Rayleigh and Double-Weibull over impaired RF/FSO system with outdated CSI," in \emph{Proc.  IEEE International Conference on Communications (ICC 2017)}, Paris, 2017, pp. 1--6.


\bibitem{prsF2}
E. Balti, M. Guizani, B. Hamdaoui, and B. Khalfi, "Aggregate hardware impairments over mixed RF/FSO relaying systems with outdated CSI," \emph{IEEE Trans. Commun.,} vol. 66, no. 3, pp. 1110--1123, Mar. 2018.



\bibitem{M1}
A. Jurado-Navas, J. M. Garrido-Balsells, J. F. Paris, and A. Puerta-Notario, "A unifying statistical model for atmospheric optical scintillation," in \emph{Numerical Simulations of Physical and Engineering Processes}, Intech, 2011.


\bibitem{M4}
H. Samimi and M. Uysal, "Performance of coherent differential phase-shift keying free-space optical communication systems in M-distributed turbulence," \emph{IEEE/OSA J. Opt. Commun. Netw.} vol. 5, no. 7, pp. 704--710, July 2013.

\bibitem{M5}
A. Jurado-Navas, J. M. Garrido-Balsells, J. F. Paris, M. Castillo-Vázquez, and A. Puerta-Notario, "Impact of pointing errors on the performance of generalized atmospheric optical channels," \emph{Opt. Express}, vol.  20, no. 11, 12550--12562, 2012.

\bibitem{M6}
I. S. Ansari, H. AlQuwaiee, E. Zedini, and M.--S. Alouini, "Information theoretical limits of free-space optical links," in \emph{Optical Wireless Communications -- An Emerging Technology}, Springer, 2016.

\bibitem{M7}
I. S. Ansari, F. Yilmaz, M.--S. Alouini, "Performance analysis of free-space optical links over M$\acute{{\rm{a}}}$laga ($ \mathcal{M} $) turbulence channels with pointing errors," \emph{IEEE Trans. Wireless Commun.}, vol. 15, no. 1, pp. 91--102, 2016.

\bibitem{prs1}
H. A. Suraweera, M. Soysa, C. Tellambura, and H. K. Garg, "Performance analysis of partial relay selection with feedback delay," \emph{IEEE Signal Process. Lett.}, vol. 17, no. 6, pp. 531-–534, Jun. 2010.

\bibitem{prs3}
M. Soysa, H. A. Suraweera, C. Tellambura, and H. K. Garg,  "Partial
and opportunistic relay selection with outdated channel estimates," \emph{IEEE Trans. Commun.}, vol. 60, no. 3, pp. 840-–850, Mar. 2012.

\bibitem{Grad}
I. S. Gradshteyn and I. M. Ryzhik, \textit{Table of Integrals, Series, and Products}. 6th ed., New York: Academic, 2000.

\bibitem{sajt}
The Wolfarm Functions Site, 2008. [Online] Available: http:/functions.wolfarm.com.

\bibitem{telfor}
M. I. Petkovic, G. T. Djordjevic,  and P. N. Ivanis, "Partial relay selection with variable gain relays and outdated CSI in mixed RF/FSO system," in \emph{Proc. 24th Telecommunications Forum (TELFOR 2016)}, Belgrade, Serbia, 2016, pp. 1--4.

\end{thebibliography}
\end{document}